# Entanglement and Applications of Pure Fourpartite of Qubit States

David Sena Oliveira and Rubens Viana Ramos

*Department of Teleinformatic Engineering, Federal University of Ceara*

*Campus do Pici, C. P. 6007, 60455-740, Fortaleza, Brazil.*

In this work we study the entanglement of pure fourpartite of qubit states. The analysis is realized through the comparison between two different entanglement measures: the Groverian entanglement measure and the residual entanglement measured with negativities. Furthemore, we discuss some applications of four-way entangled fourpartite states.

## 1. Introduction

Quantum entanglement is the key property that allows the design of powerful quantum communication protocols and quantum algorithms. Hence, the analysis of entanglement through its classification and quantification is a crucial task in quantum information. Well established entanglement measures for two qubit states (pure and mixed) [1-4] and pure three qubit states [5-7] have been reported. Considering multi-qubit states with $n>3$, some entanglement measures have been considered [8] and two interesting are the Groverian entanglement measure [9-11] and the generalization of the residual entanglement based on the negativity proposed in [7]. In this work we use these last two measures to study the entanglement of some pure fourpartite of qubit states having four-way entanglement, including graphs states [12,13]. Furthermore, we present some applications of fourpartite states in teleportation and quantum communication.

## 2. Entanglement Measures for Pure Fourpartite of Qubit States

In this work we are concerned only with the Groverian entanglement measure and the residual entanglement measured with negativities. Thus, in this section we give a brief review of them.

The Groverian entanglement is an operational measure based on the quantum search algorithm proposed by Grover [9-11]. Basically, it relates the entanglement of the quantum state that represents the database with the average probability of finding a marked state. The lowest the entanglement of the input state the largest is the probability of finding the marked state. In order to maximize the probability of the quantum search to find the marked state, local unitary operation are allowed. In this way, for an input state $|\psi\rangle$, the (average) success probability of the quantum search is

$$P_{\max}(\psi) = \max_{U_1,\cdots,U_n} \frac{1}{N} \sum_{m=0}^{N-1} \left| \langle m | (U_G)^k (U_1 \otimes \cdots \otimes U_n) | \psi \rangle \right|^2, \qquad (1)$$

where $U_G$ is the Grover operator that is applied $k$ times, $|m\rangle$ is the marked state, $N=2^n$ and $n$ is the number of qubits, and $U_1,\ldots,U_n$ are the allowed local unitary operations that can be changed in order to maximize the success probability. Since $\langle m|(U_G)^k(U_1\otimes\ldots\otimes U_n)$ is a pure disentangled state, equation (1) can be rewritten as

$$P_{\max}(\psi) = \max_{|e_1,\ldots,e_n\rangle} |\langle e_1,\ldots,e_n|\psi\rangle|^2 \qquad (2)$$

In (2), $|e_i\rangle$ is a pure one qubit state. At last, the Groverian entanglement measure is given by

$$E_G(\psi) = \sqrt{1 - P_{\max}(\psi)} \qquad (3)$$

As can be seen in (2)-(3), the Groverian entanglement measure consists in finding the closest disentangled state (formed by the tensor product of single-qubit states) of the state whose entanglement one wishes to measure. The distance measure used is the fidelity. Considering fourpartite states of qubits, one has

$$P_{\max}(\psi) = \max_{|e_1,\ldots,e_n\rangle} \left| \begin{bmatrix} (\langle 0|\cos(\theta_1) + \langle 1|e^{i\varphi_1}\sin(\theta_1)) \otimes (\langle 0|\cos(\theta_2) + \langle 1|e^{i\varphi_2}\sin(\theta_2)) \otimes \\ (\langle 0|\cos(\theta_3) + \langle 1|e^{i\varphi_3}\sin(\theta_3)) \otimes (\langle 0|\cos(\theta_4) + \langle 1|e^{i\varphi_4}\sin(\theta_4)) \end{bmatrix} |\psi\rangle \right|^2 \qquad (4)$$

and the maximization is taken over the angles $\theta_i$ and $\varphi_i$: $\partial P/\partial\theta_i = \partial P/\partial\varphi_j = 0$ for all $i,j=1,2,3,4$. Here, instead of looking for analytical equations for $P_{max}$ and $E_G$, as it was done in [10,11], we developed a genetic algorithm that can find $P_{max}$ and $E_G$ for any number of qubits. For example, for the W state with $n$ qubits

$$|W_n\rangle = \frac{1}{\sqrt{2^n}}(|00\ldots01\rangle + |00\ldots10\rangle + \ldots + |01\ldots00\rangle + |10\ldots00\rangle) \qquad (5)$$

the Groverian entanglement can be analytically calculated by the equation [9]

$$E_G(W_n) = \sqrt{1 - \left(\frac{n-1}{n}\right)^{n-1}} \qquad (6)$$

The comparison between the analytical result given by (6) and the numerical result achieved by our genetic algorithm, can be seen in Fig. 1

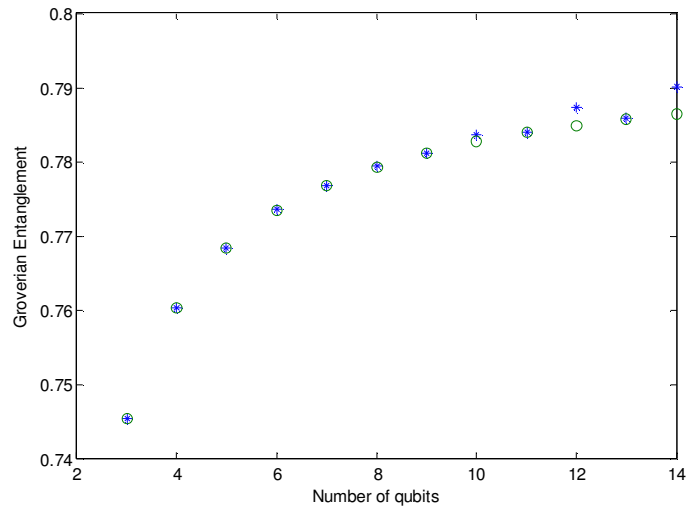

Figure 1 – Groverian entanglement for W states having 3-14 qubits. Analytical (o) and nuemrical (*) results.

In Fig. 2 one can see the Groverian entanglement of the state $a|0000\rangle+b|1111\rangle$, obtained numerically. Its maximum occurs when $a=b=1/\text{sqrt}(2)$, $E_G=0.707$.

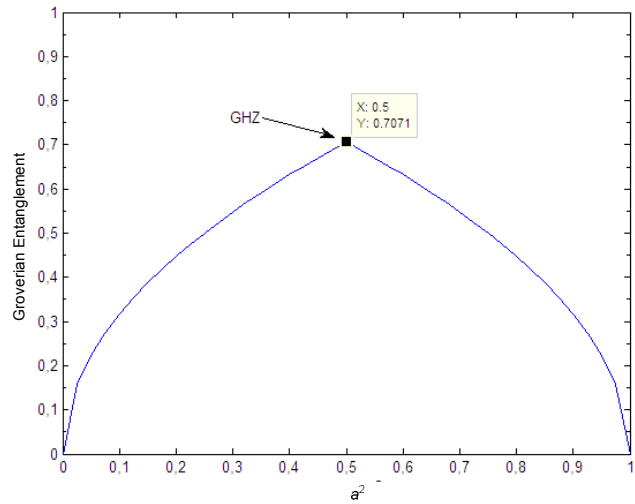

Figure 2 – Groverian entanglement for the state $a|0000\rangle+b|1111\rangle$ versus $a^2$.

The second entangled measure to be considered is the residual entanglement. It was firstly proposed in [6] and modified and generalized in [7], where instead of using the concurrence for measuring the bipartite entanglements, the negativity was employed. Considering pure tripartite of qubit states, the three-way entanglement can be measured by the residual entanglement measure $\pi_3$ defined in [7] as:

$$\pi_3 = \frac{1}{3}(\pi_A + \pi_B + \pi_C) \quad (7)$$

$$\pi_A = N^2_{A\_BC} - N^2_{AB} - N^2_{AC}, \quad (8)$$

$$\pi_B = N^2_{B\_AC} - N^2_{AB} - N^2_{BC} \quad (9)$$

$$\pi_C = N^2_{C\_AB} - N^2_{AC} - N^2_{BC} \quad (10)$$

where, for example, $N^2_{A\_BC} = C^2_{A\_BC} = 2(1 - Tr\rho_A^2)$ is the negativity of the system composed by the single subsystem A and the bipartite system BC, while $N_{AB} = Tr\sqrt{\rho_{AB}^{T_A}(\rho_{AB}^{T_A})^\dagger} - 1$ is the (normalized) negativity of the subsystems AB [4]. In general $\pi_A \neq \pi_B \neq \pi_C$ and, for W-class states, $\pi_3$ maybe larger than zero. The extension for fourpartite states is straightforward

$$\pi_4 = \sqrt[4]{\pi_A \pi_B \pi_C \pi_D} \quad (11)$$

$$\pi_A = N^2_{A\_BCD} - N^2_{AB} - N^2_{AC} - N^2_{AD} \quad (12)$$

$$\pi_B = N^2_{B\_ACD} - N^2_{AB} - N^2_{BC} - N^2_{BD}. \quad (13)$$

$$\pi_C = N^2_{C\_ABD} - N^2_{AC} - N^2_{BC} - N^2_{CD} \quad (14)$$

$$\pi_D = N^2_{D\_ABC} - N^2_{AD} - N^2_{BD} - N^2_{CD} \quad (15)$$

Here, we use the geometric mean instead of arithmetic mean because this last one is not zero for fourpartite states formed by the tensor product of a single-qubit and a three-way entangled tripartite state. As an example, let us consider the state [12]

$$|\chi\rangle = \frac{|\xi_0\rangle + |\xi_1\rangle}{\sqrt{2}} \quad (16)$$

$$|\xi_0\rangle = \cos(\theta)|0000\rangle - \sin(\theta)|0011\rangle - \sin(\phi)|0101\rangle + \cos(\phi)|0110\rangle. \quad (17)$$

$$|\xi_1\rangle = \cos(\phi)|1001\rangle + \sin(\phi)|1010\rangle + \sin(\theta)|1100\rangle + \cos(\theta)|1111\rangle \quad (18)$$

The four-way entanglement measured by $\pi_4$ versus the angles $\theta$ and $\phi$ can be seen in Fig. 3.

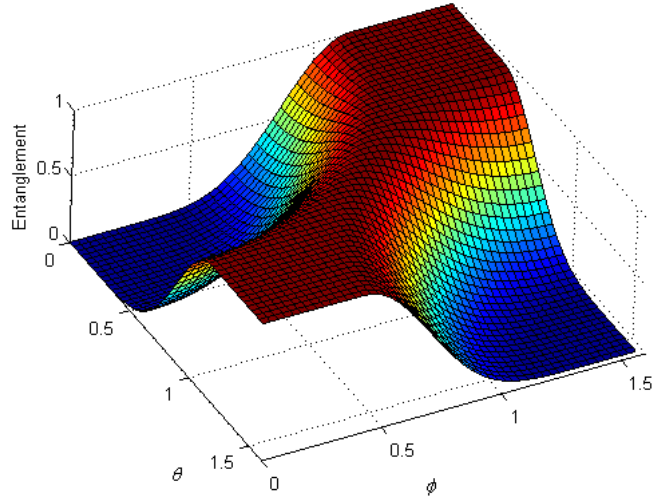

Figure 3 – Entanglement of $\chi$, Eq. (16)-(18), versus $\theta$ and $\phi$, measured by $\pi_4$.

In Table 1 one can see the entanglement of several fourpartite of qubit states measured by $E_G$ and $\pi_4$.

| Quantum State | $E_G$ | $\pi_4$ |
|---|---|---|
| $\|\psi_1\rangle=(\|01\rangle+\|10\rangle)/2^{1/2}\otimes(\|0\rangle+\|1\rangle)/2^{1/2}\otimes(\|0\rangle+\|1\rangle)/2^{1/2}$ | 0.707 | 0 |
| $\|\psi_2\rangle=(\|00\rangle+\|11\rangle)/2^{1/2}\otimes(\|00\rangle+\|11\rangle)/2^{1/2}$ | 0.866 | 0 |
| $\|\psi_3\rangle=(\|000\rangle+\|111\rangle)/2^{1/2}\otimes(\|0\rangle+\|1\rangle)/2^{1/2}$ | 0.707 | 0 |
| $\|\psi_4\rangle=(\|001\rangle+\|010\rangle+\|100\rangle)/3^{1/2}\otimes(\|0\rangle+\|1\rangle)/2^{1/2}$ | 0.745 | 0 |
| $\|\xi_0\rangle=(\|0000\rangle-\|0011\rangle-\|0101\rangle+\|0110\rangle)/2$ | 0.707 | 0 |
| $\|\xi_1\rangle=(\|1001\rangle+\|1010\rangle+\|1100\rangle+\|1111\rangle)/2$ | 0.707 | 0 |
| $\|\psi_5\rangle\equiv\|\chi^{00}\rangle=(\|\xi_0\rangle+\|\xi_1\rangle)/2^{1/2}$ | 0.866 | 1 |
| $\|\psi_6\rangle=(\|0000\rangle+\|1111\rangle)/2^{1/2}$ | 0.707 | 1 |
| $\|\psi_7\rangle=(\|0001\rangle+\|0010\rangle+\|0100\rangle+\|1000\rangle)/2$ | 0.76 | 0.14903 |
| $\|\psi_8\rangle=(\|0000\rangle+\|0101\rangle+\|1000\rangle+\|1110\rangle)/2$ | 0.707 | 0.25993 |
| $\|\psi_9\rangle=(\|0000\rangle+\|1011\rangle+\|1101\rangle+\|1110\rangle)/2$ | 0.81 | 0.75 |
| $\|\psi_{10}\rangle=(\|0001\rangle+\|0110\rangle+\|1000\rangle)/3^{1/2}$ | 0.81 | 0.40861 |
| $\|\psi_8\rangle=(\|0000\rangle+\|0111\rangle+\|1011\rangle+\|1100\rangle)/2$ | 0.866 | 1 |
| $\|\psi_9\rangle=(\|0000\rangle-\|0101\rangle+\|1010\rangle+\|1111\rangle)/2$ | 0.866 | 1 |

Table 1 – Entanglement of several fourpartite of qubit states measured with $E_G$ and $\pi_4$.

As can be seen in Table 1, there are fourpartite states that have no four-way entanglement ($\|\psi_1\rangle,…,\|\psi_4\rangle,\|\xi_0\rangle,\|\xi_1\rangle$) but their $E_G$ is larger than zero. This happens because the Groverian entanglement

detects the presence of any type of entanglement. For example, the state $|\psi_1\rangle$ has a bipartite entanglement and, hence, $|\psi_1\rangle$ can not be written as a tensor product of four single-qubits states, thus, $P_{max}$ will be lower than one what makes $E_G$ larger than zero. It is interesting to observe that the states $|\psi_6\rangle$, $|\psi_7\rangle$, $|\psi_8\rangle$, $|\psi_9\rangle$ and $|\psi_{10}\rangle$ correspond to different classes of entanglement [14]. This can be detected by $\pi_4$ but not by $E_G$. Since the Groverian entanglement measure is not a reliable measure for genuine fourpartite entanglement, hereafter we are going to use only the residual entanglement $\pi_4$.

Now, let us consider the following graph states having four qubits [13]

$$|G\rangle = U_{12}^{b_0} U_{13}^{b_1} U_{14}^{b_2} U_{23}^{b_3} U_{24}^{b_4} U_{34}^{b_5} \left(\frac{|0\rangle+|1\rangle}{\sqrt{2}}\right)^{\otimes 4} \quad (19)$$

$$U_{ij} = |0\rangle\langle 0|_i \otimes I_j + |1\rangle\langle 1|_i \otimes Z_j \quad (20)$$

In (19) $U_{ij}$ is a controlled-phase gate applied on the qubits $i$ and $j$, and $b_k \in \{0,1\}$ ($k=0,\ldots,5$). It can be checked (there are only 64 possibilities) that a state in (19) is maximally entangled, $\pi_4=1$, if the graph that represents the state (for example, there is an edge between vertices 1 and 2 if $b_0=1$) is completely connected, otherwise the state has not genuine fourpartite entanglement, $\pi_4=0$. This has been pointed out in [15]. Some examples can be seen in Fig. 4.

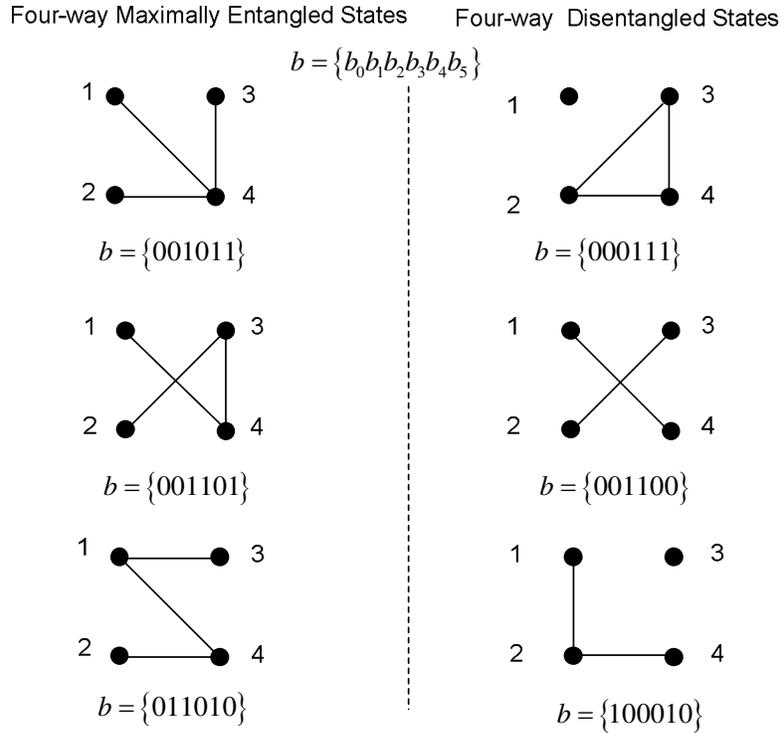

Figure 4 – Maximally entangled and disentanglement graph states according to $\pi_4$.

## 3. Applications of Four-Way Entangled Pure Fourpartite of Qubit States

Fourpartite state having four-way entanglement are particularly useful for teleportation of two qubit gates. For example, the state $(|0000\rangle+|0111\rangle+|1011\rangle+|1100\rangle)/2$ has been used for teleportation of the CNOT operation [16]. The state $(|0000\rangle-|0101\rangle+|1010\rangle+|1111\rangle)/2$, by its turn, can be used for teleportation of the two qubit quantum circuit shown in Fig. 5.

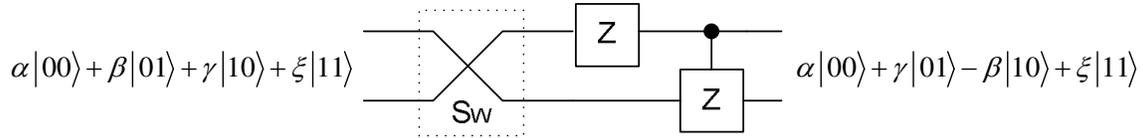

Figure 5 – Quantum circuit to be teleported using the fourpartite state $(|0000\rangle-|0101\rangle+|1010\rangle+|1111\rangle)/2$. $S_W$ is the swap gate.

The quantum circuit for teleportation of the circuit in Fig. 5 can be seen in Fig. 6.

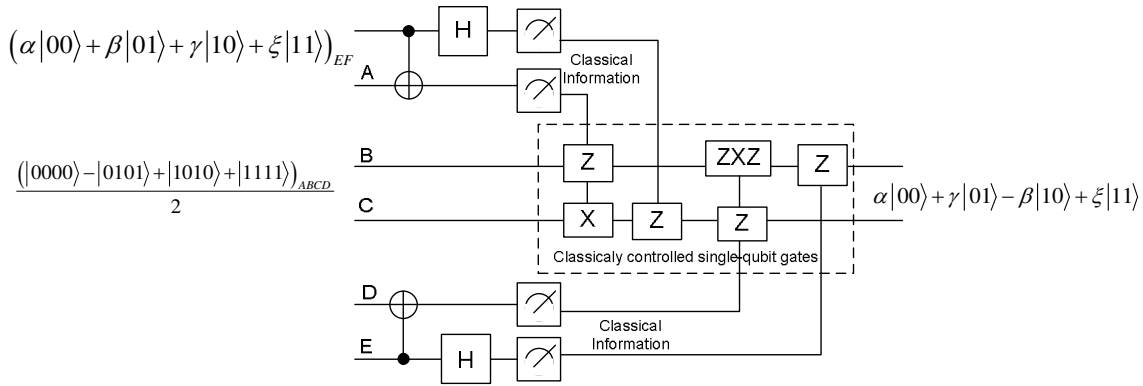

Figure 6 – Quantum circuit for teleportation of the quantum circuit in Fig. 5.

The error corrections to be applied according to the results of the measurements are shown in Table 2.

| Measurement Results (AFDE) | Quantum Operation | Measurement Results (AFDE) | Quantum Operation |
|---|---|---|---|
| 0000 | $I \otimes I$ | 1000 | $Z \otimes X$ |
| 0001 | $Z \otimes I$ | 1001 | $I \otimes X$ |
| 0100 | $I \otimes Z$ | 1100 | $Z \otimes ZX$ |
| 0101 | $Z \otimes Z$ | 1101 | $I \otimes ZX$ |
| 0010 | $ZXZ \otimes Z$ | 1010 | $ZX \otimes ZX$ |
| 0011 | XZ⊗Z | 1011 | X⊗ZX |
| 0110 | ZXZ⊗I | 1110 | ZX⊗X |
| AB0111 | XZ⊗I | 1111 | X⊗X |

Table 2 – Error correction table for teleportation scheme shown in Fig. 6.

As a second application, let us assume that each qubit of the quantum state $(|00\rangle_{AB}+|11\rangle_{AB})/2^{1/2}$ is sent through a noisy channel, modeled by the interaction with environment through the unitary operations $U_A$ for qubit $A$ and $U_B$ for qubit $B$. This scheme is shown in Fig. 7.

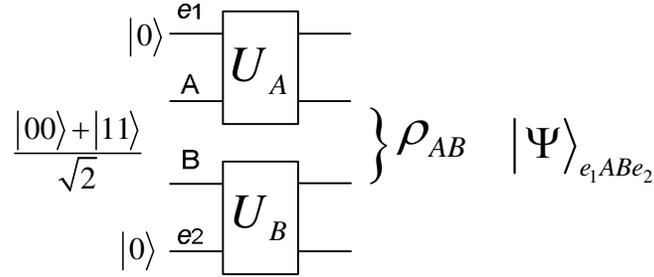

Figure 7 – Two entangled qubits sent through noisy channels modeled by unitary operations $U_A$ and $U_B$.

Assuming $U_A=\exp(i\theta X\otimes X)$ and $U_B=\exp(i\phi X\otimes X)$, and that the initial state of the environment is $|0\rangle$ for both noisy channels, the entanglement of the total fourpartite state $|\Psi\rangle_{e_1ABe_2}$ and the entanglement of the bipartite state $\rho_{AB}$ at the channel's output can be calculated. The quantum states are given by

$$|\Psi\rangle_{e_1ABe_2} = (U_1 \otimes U_2)|0\rangle_{e_1}\left(\frac{|00\rangle+|11\rangle}{\sqrt{2}}\right)_{AB}|0\rangle_{e_2} \quad (21)$$

$$\rho_{AB} = Tr_{e_1e_2}\left(|\Psi\rangle_{e_1ABe_2}\right) \quad (22)$$

In Fig. 8 one can see the entanglement $\pi_4\left(|\Psi\rangle_{e_1ABe_2}\right)$ and the Vidal-Werner negativity of $\rho_{AB}$, $N(\rho_{AB})$, versus the angles $\theta$ and $\phi$. As expected, the larger $\pi_4$ the lower is $N$ and vice-versa.

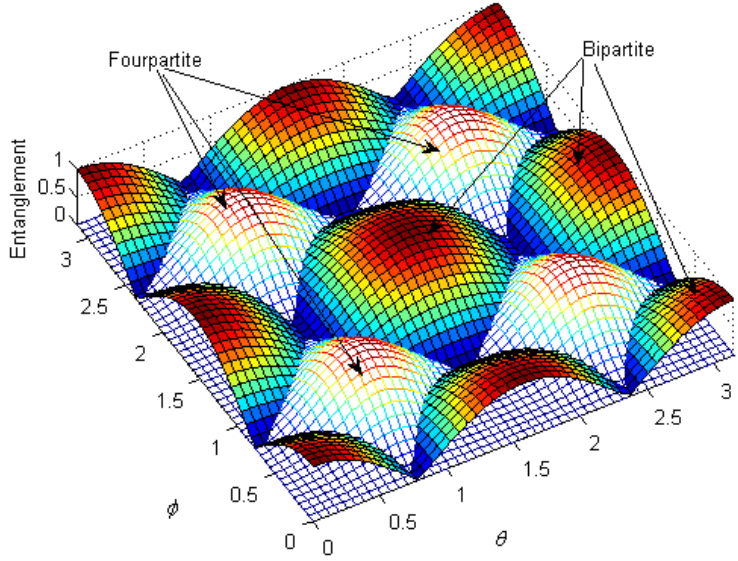

Figure 8 – Entanglements of the fourpartite and bipartite states shown in Fig. 7.

Now, following [12,13], lets us consider the following maximally entangled ($\pi_4$=1) quantum state

$$|\xi_0\rangle = \frac{1}{2}\left[|0001\rangle + |0010\rangle + |0100\rangle + |0111\rangle\right] \quad (23)$$

$$|\xi_1\rangle = \frac{1}{2}\left[|1000\rangle + |1011\rangle + |1101\rangle + |1110\rangle\right] \quad (24)$$

$$|\chi^{11}\rangle = \frac{1}{\sqrt{2}}\left(|\xi_0\rangle + |\xi_1\rangle\right) \quad (25)$$

This state is locally equivalent to a maximally entangled graph. According to (19), one has

$$|\chi^{11}\rangle = \left(H_1 \otimes I_2 \otimes I_3 \otimes HZH_4\right) U_{12} U_{13} U_{14} \left(\frac{|0\rangle + |1\rangle}{\sqrt{2}}\right)^{\otimes 4} \quad (26)$$

Since graph states are also stabilizers states, the state $|\chi^{11}\rangle$ is also a stabilizer state and it is stabilized by

$$S = X^a \otimes X^b \otimes X^c \otimes X^d \quad (27)$$

where $a$, $b$, $c$ and $d$ form a binary string with even number of 1's. Hence, it is robust against bit flip at any two qubits or bit flip at all four qubits. The state $|\chi^{11}\rangle$ is also LU equivalent to ($|0000\rangle$-

$|0111\rangle+|1000\rangle+|1111\rangle)/2,$

$$\left|\chi^{11}\right\rangle = \left(I_1 \otimes H_2 \otimes H_3 \otimes H_4\right)\frac{\left[|0000\rangle - |0111\rangle + |1000\rangle + |1111\rangle\right]}{2} \quad (28)$$

Using the quantum circuit in Fig. 9, the state $(|0000\rangle-|0111\rangle+|1000\rangle+|1111\rangle)/2$ can be employed for quantum teleportation of the bipartite state $\alpha|00\rangle+\beta|11\rangle$.

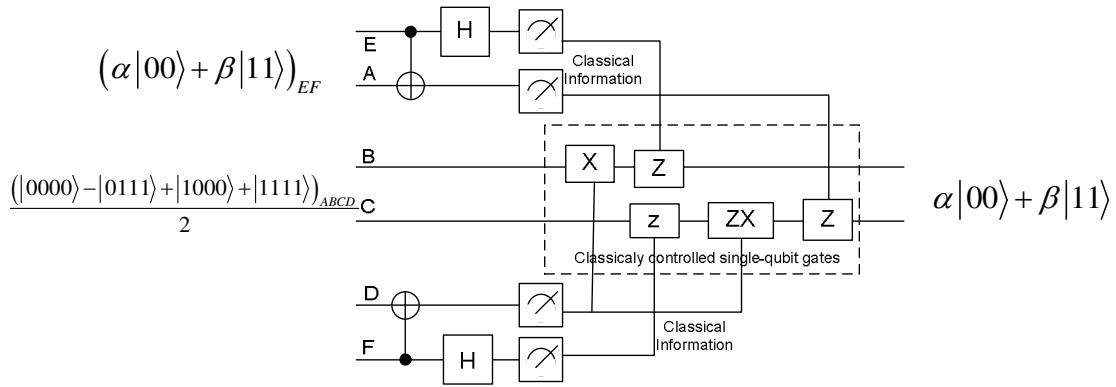

Figure 9 – Quantum circuit for teleportation of the quantum state $\alpha|00\rangle+\beta 11\rangle$.

At last, let us assume a network communication with three parts, Alice, Bob and Charlie. They will run a protocol in which Bob and Charlie want to exchange quantum bits by quantum teleportation. Alice, by its turn, play the role of a TTP (trusted third part), this means that Bob and Charlie believe that Alice is 100% honest. Let us suppose that, for some reason, Bob (Charlie) sends quantum states to Charlie (Bob) and Charlie (Bob) cannot know its "value" before some amount of time. We stress that our goal here is not to discuss all details concerning this protocol, but only to show that the fourpartite state $|\chi^{00}\rangle$ ($|\psi_5\rangle$ in Table 1) can be useful in a quantum communication protocol. The following protocol with fourpartite states can be used to realize the task

1. Alice produces states $|\chi^{00}\rangle$ ($|\psi_5\rangle$ in Table 1), keeps qubits A and B with her and she sends the qubit C to Bob and the qubit D to Charlie.
2. After Alice measuring her qubits, Bob and Charlie will share a maximally entangled two qubit states, according to Table 3. For example, if her measurements result is $|00\rangle$, then Bob and Charlie will share the state $(|00\rangle-|11\rangle)/2^{1/2}$, that can be used for quantum teleportation.
3. Having the entangled qubits, Bob and Charlie can initiate a quantum teleportation, but they cannot complete it because they do not know which Bell state is being used. Hence, they have to wait for Alice' information before conclude the teleportation.

4. Alice informs to Bob and Charlie the results of the measurements in qubits A and B. Knowing this information Bob (Charlie) informs to Charlie (Bob) which single-qubit operations he has to apply in order to obtain the state teleported.

| Measurement Result | $|\psi_{CD}\rangle$ |
|---|---|
| $|00\rangle$ | $(|00\rangle-|11\rangle)/2^{1/2}$ |
| $|01\rangle$ | $(|10\rangle-|01\rangle)/2^{1/2}$ |
| $|10\rangle$ | $(|10\rangle+|01\rangle)/2^{1/2}$ |
| $|11\rangle$ | $(|00\rangle+|11\rangle)/2^{1/2}$ |

Table 3 – State shared by Bob and Charlie according to Alice's measurement results.

## 4. Conclusions

We have firstly compared two entanglement measures for fourpartite of qubit states: the Groverian entanglement and the residual entanglement based on negativity. The entanglements of several fourpartite states were calculated and the residual entanglement is considered more appropriate for the entanglement calculation. However, differently from initially proposed in the literature, we use the geometric mean instead of arithmetic mean because this last one is not zero for fourpartite states formed by the tensor product of a single-qubit and a three-way entangled tripartite state. Following, we described some applications of entangled fourpartite states. Firstly, we showed a quantum circuit for teleportation of a two-qubit quantum circuit composed by a swap, Z and controlled-Z gates. The quantum state required for this is $(|0000\rangle-|0101\rangle+|1010\rangle+|1111\rangle)/2$. After, we showed the entanglement change of an initially maximally entangled bipartite state that interacts with two different environments, modeled by the unitary operations $U_A=\exp(i\theta X\otimes X)$ and $U_B=\exp(i\phi X\otimes X)$. As expected, when the total (fourpartite) entanglement of the bipartite state with the environments is maximal, the entanglement of the bipartite state is minimal. After, we considered the state $|\chi^{11}\rangle$ (eqs. (23)-(25)) and we showed that it is locally equivalent to the graph state $U_{12}U_{13}U_{14}[(|0\rangle+|1\rangle)/2^{1/2}]^{\otimes 4}$ and locally equivalent to $(|0000\rangle-|0111\rangle+|1000\rangle+|1111\rangle)/2$, which is used for the teleportation of the bipartite state $\alpha|00\rangle+\beta|11\rangle$, using the scheme in Fig. 9. At last, we presented a quantum communication protocol using the fourpartite states $|\chi^{00}\rangle$. In this three-part protocol, Bob and Charlie can exchange a quantum bit by using quantum teleportation only if Alice permits.

## Acknowledgements
This work was supported by the Brazilian agencies FUNCAP and CAPES.